\documentstyle[draft,psfig]{mn}
\newcommand{\mnref}[1]{\hangindent=0.5in \hangafter=1 #1 \par}

\newcommand{\mn}{MNRAS}

\newcommand{\apj}{ApJ}
\newcommand{\apjs}{ApJS}
\newcommand{\aaa}{A\&A}
\newcommand{\aas}{A\&AS}
\newcommand{\gtn}{\mbox{G29.96$-$0.02}}
\newcommand{\etal}{et al.\ }
\newcommand{\kms}{kms$^{-1}$}
\newcommand{\Msolar}{\mbox{\,$\rm M_{\odot}$}}
\newcommand{\brg}{\mbox{Br$\gamma$}}

\title{Molecular and Ionised Gas Motions in the Compact HII region
G29.96--0.02} 
\author[S.L. Lumsden \& M.G. Hoare]
{S.L. Lumsden$^{1}$ and M.G. Hoare$^{2}$\\
{}$^1$ {\em Anglo-Australian Observatory, PO Box 296, Epping, NSW 1710,
Australia -- sll@aaoepp.aao.gov.au}\\
{}$^2$ {\em Department of Physics and Astronomy,
University of Leeds, Leeds LS2 9JT, UK -- mgh@ast.leeds.ac.uk}}

\begin{document}
%\label{firstpage}

\maketitle

\begin{abstract}
We present a new observation of the compact HII region, \gtn, that allows us to
compare the velocity structure in the ionised gas and surrounding molecular gas
directly.  This allows us to remove most of the remaining ambiguity about the
nature of this source.  In particular, the comparison of the velocity structure
present in the 4$^{1,3}$S--3$^{1,3}$P HeI lines with that found in the 1--0
S(1) H$_2$ line convincingly rules out a bow shock as being important to the
kinematics of this source.  Our new observation therefore agrees with our
previous conclusion, drawn from a velocity resolved HI Br$\gamma$ map, that
most of the velocity structure in \gtn\ can largely be explained as a result of
a champagne flow model.  We also find that the best simple model must invoke a
powerful stellar wind to evacuate the `head' of the cometary HII region of
ionised gas.  However, residual differences between model and data tend to
indicate that no single simple model can adequately explain all the observed
features.

\end{abstract}


\section{Introduction}
The study of ultracompact (UC) H~II regions has been stimulated in recent years
by the sparsely sampled VLA survey of Wood \& Churchwell (1989).  One of the
greatest puzzles of the Wood \& Churchwell survey is the total number of such
regions found.  By taking both the total number of O stars in the galaxy
(simply from extrapolating the number of optically visible OB associations),
and their lifetimes on the main sequence, they predicted the total number of
such sources that might still be expected to be in the UC H~II phase.
Comparing this with the numbers actually found led Wood \& Churchwell to
conclude that there was an order of magnitude more sources in their VLA survey
than there should have been. Since there is no other evidence for the massive
star formation rate being this much larger than expected, they considered that
the likeliest solution to this problem is that some mechanism is constraining
the outward expansion of the H~II region.  Wood \& Churchwell estimate that the
UC H~II phase must last about ten times longer than the $\sim10^4$ years
predicted on the basis of simple Str\"{o}mgren sphere expansion.  Another key
result from the Wood \& Churchwell study is that at least 20\% of these regions
have a `cometary' appearance.  A subsequent survey by Kurtz \etal (1994) and a
VLA survey of much larger, optically visible, H~II regions by Fich (1993) also
found a similar proportion of this type.

These discoveries stimulated interest in the study of new models for the
evolution of compact HII regions.  Some of the key models that have been
developed recently that can explain the lifetime of the compact HII regions can
be summarised as follows:\\

\hspace{3mm}$\bullet$ Bow shock models: The near parabolic shape of many of
these sources lead Wood \& Churchwell to develop a bow-shock interpretation for
UC H~II regions which provides an explanation for both of the lifetime and the
cometary morphologies.  The basis of this model is that an OB star moves
supersonically through a molecular cloud ($v_*>0.2$~\kms) and the stellar wind
from the star supports a bow shock along its direction of motion (Mac-Low et
al.\ 1991, Van Buren \& Mac-Low 1992).  This shock can trap the ionisation
front (IF), preventing it from expanding and the lifetime of the UC H~II phase
in this model is simply the star's crossing time through the cloud, typically
of the order of 10$^5$ years.

\hspace{3mm}$\bullet$ Mass loading models: Dyson and co-workers (Dyson et al.\
1995; Redman et al.\ 1996) have investigated the interaction of a stellar wind
and a clumpy molecular cloud material.  This provides a source of fresh neutral
material to soak up ionising photons which is an efficient mechanism for
slowing the expansion. A density gradient of mass-loading clumps can also
account for the cometary regions (Williams et al.\ 1996, Redman et al.\ 1998).
Lizano et al.\ (1996) offer a slightly different perspective on the same
process.  Hollenbach et al.\ (1994) invoked photoevaporation of a disk around
the exciting star itself to feed material, but this does not help to account
for the known morphologies.

\hspace{3mm}$\bullet$ `Environmental' models: De Pree, Rodriguez \& Goss (1995)
suggested that there was evidence that the ambient density in the natal
molecular clouds was much higher than assumed in traditional Str\"{o}mgren
sphere expansion arguments.  In a sense this `model' argues that there is in
fact no problem, rather that Wood \& Churchwell used inaccurate parameters in
calculating the expected growth rates of the HII regions initially.  In this
picture the likeliest explanation for the observed fraction of compact HII
regions that appear `cometary' is given by the champagne flow model described
next.

\hspace{3mm}$\bullet$ Champagne flow (blister) models: HII regions with
cometary morphology were originally labelled as `blisters' by Israel (1978)
because of their propensity for being found near the edges of molecular clouds
(the classical example being the Orion nebula itself of course).  Icke, Gatley
\& Israel (1980) noted that if there is a density gradient in the ambient gas
the H~II region will expand fastest in the low density direction and so become
very asymmetric. Tenorio-Tagle and co-workers in a series of papers (see Yorke
et al. 1983 for a summary) developed numerical models of the evolution of HII
regions for this asymmetric expansion.  They found that the pressure gradient
set up when the IF reaches the edge of the cloud causes a `champagne' flow of
ionized gas away from the cloud with velocities of order 30 \kms.  Although in
its original form this model does not answer the lifetime problem, it may when
combined with either higher ambient densities, stellar wind trapping of the
ionization front or mass loading.

Whilst it is clear from the above that no single observational test can
absolutely prove that one of these models is correct, the velocity structure
within the ionised and molecular gas provides a key discriminating test.
Previous attempts to study the molecular gas in the mm regime have turned up
unexpected results. Hot ammonia clumps (Cesaroni et al. 1994) and water masers
(Hofner \& Churchwell 1996) often appear to be more associated with a
subsequent generation of star formation rather than probing the adjacent
molecular gas. Observations of tracers of the more quiescent gas such as CO
have mostly been made with low spatial resolution and optical depth and
excitation effects can also make interpretation difficult.

G29.96--0.02 is perhaps the best studied of the compact cometary sources for
which evidence of a bow shock has been found.  Wood \& Churchwell (1991)
carried out high resolution radio recombination line studies, and the results
were analysed by Van Buren \& Mac Low (1992), who claim good agreement with
their bow-shock model.  We showed in a previous paper (Lumsden \& Hoare 1996:
hereafter LH96) however, that their conclusions were biased because of the low
sensitivity to extended structure in their data.  By mapping the HI Br$\gamma$
line we instead found much better agreement with a model that had a strong
champagne flow component.  In this paper we present new data, allowing us to
tie the motions in the molecular and ionised gas together without ambiguity for
the first time.

\section{Observations}
We obtained high resolution ($R\sim20000$) spectra of the 1--0 S(1) H$_2$ line
with the common user IR array spectrometer CGS4 on UKIRT on the night of 4 July
1997.  We used the echelle grating within CGS4, with a one pixel wide slit
(each pixel corresponding to 1.0$''$ in the dispersion direction and 1.5$''$
along the slit).  The effective resolution of this combination was measured to
be $18\pm1$~\kms\ on several bright OH night sky lines (intrinsic width assumed
to be $\ll10$\kms), and from a krypton arc lamp.  To fully sample the
resolution element, the array was stepped six times across two pixels.  This
oversampling is extremely useful in determining accurate line profiles for our
data.  The slit was set at a position angle of 60$^\circ$, in order to be
approximately along the symmetry axis of \gtn.

Our basic observing technique was to observe at each position for ten minutes
(composed of five two minute exposures), and then take a separate two minute
sky frame.  Since the night sky lines are well separated from the emission
lines in our target we do not need to achieve a good cancellation of these.
The separate sky frame acts as a dark frame to remove dark current from the
array more than to subtract the night sky emission.  We coadded five such data
blocks (ie total time per position on source was 50 minutes).  We observed
three positions in total, aligned with the peak flux from the object and
6 arcseconds northwest and southeast of this.  Our data can be compared
directly with the equivalent HI Br$\gamma$ data from LH96.

The most crucial part of the data reduction is accurate wavelength calibration
since we wish to compare the relative velocities of the observed lines
directly.  Unfortunately, flexure can give rise to significant wavelength
shifts in CGS4, so it is not feasible to use arc lamps to calibrate the
wavelength scale.  Instead, we derived a wavelength calibration using the
actual night sky lines on the frame.  This ensures that we are not affected by
the flexure problems, since the calibration is now self consistent with the
emission line data: it does however mean that the absolute reference for the
velocity is only as accurate as the night sky emission lines.  Each frame was
shifted onto a constant wavelength scale before the co-addition into the final
three images.  In addition to the correction for flexure (which appears as
a bulk wavelength shift for all slit positions), the projection of the slit
onto the array is curved.  We also corrected for this curvature.  We tested
for residual errors in these corrections by measuring wavelengths of night
sky lines in different frames and at different positions along the slit.
We are confident that the residual errors in the measured relative velocities
due to these corrections are $<1$kms$^{-1}$.

For the OH lines, we used the tabulated wavelengths given in Oliva \& Origlia
(1992), and allowed for the splitting of the doublets (since this leads to
resolvable lines in some instances).  With the 256$\times$256 array now used in
CGS4, the wavelength coverage ($\sim0.027\mu$m), allowed us to observe 11
identifiable lines.  We were able to derive a very good fit to the tabulated
values, with a final rms error of $1.1\times10^{-5}\mu$m (or only 1.5kms$^{-1}$
in velocity terms).  The best fit to CGS4 echelle data is quadratic in the
estimated wavelength (note that this is different to our assumption in LH96
where we adopted a linear shift from the brightest arc or OH lines).  Lastly,
we corrected the observed velocities into standard v$_{LSR}$ values by
correcting for the solar motion.  We checked the derived
dispersion correction for the spectrum using this method with that obtained
from a krypton arc lamp.  Although there was an offset because of flexure
as already noted, the actual dispersion correction was very similar, giving
us confidence in our adopted approach.

We now return briefly to the case of the absolute wavelength calibration of the
Br$\gamma$ data published in LH96.  As part of a separate project we have
obtained spectra of planetary nebulae (PN) with the same configuration. Of
these $\sim10$ are known to have accurate measured radial velocities.  We used
this information to derive a calibration from the OH night sky lines at
Br$\gamma$ as well (it was possible to apply this method since the PN data were
also obtained with the larger array, whereas the original LH96 data were
obtained using a smaller 58$\times$62 device).  Although the tabulated Oliva \&
Origlia values again gave a good dispersion correction, we found an offset
between the PN data and the known radial velocities.  We derived an average
correction to apply, and then used this knowledge to derive a similar
wavelength calibration for \gtn.  The result is rather different to that given
in LH96.  For the pixel with the largest flux, we derive a velocity of
97.2kms$^{-1}$.  We also carried out observations of a smaller set of PN to
calibrate the error in the H$_2$ data.  From this we derive the velocity of the
pixel with the peak HeI intensity as 98.3kms$^{-1}$, in very good agreement
with the HI line value.  We also used the same PN data to estimate the relative
error in the velocity scale between the HeI lines and H$_2$ data, and found an
average difference of $\sim2$kms$^{-1}$ assuming the H$_2$ and HeI trace the
same structure in the PN.  Therefore we assume that the overall scale of both
datasets is accurate to $\pm3$kms$^{-1}$ (which is also consistent with the
derived errors in the actual arc fitting process for both datasets as well).
We are however confident that there is no evidence for any difference in the
absolute velocity of the HI and HeI from our data, and have therefore assumed
that they are the same in what follows (ie we have normalized the data to have
the same velocity for the point with the largest observed line flux in the
ionised component).

Figure 1 shows the coadded spectrum of the region around the peak of the
ionised emission.  The lines evident are the 4$^{3,1}$S--3$^{3,1}$P transitions
of HeI, and the 1--0 S(1) transition of H$_2$.  In comparing our data with the
models we also require very accurate theoretical or measured wavelengths.  The
vacuum rest wavelength for the H$_2$ line was taken as 2.121833$\mu$m from
Bragg, Brault \& Smith (1982), whilst the values for the HeI lines were derived
from energy levels in Bashkin \& Stoner (1975) from which we found
2.113772$\mu$m and 2.112743/2.112657/2.112584$\mu$m for the singlet and triplet
transitions respectively.  For the singlet transition there is no ambiguity as
to which spin state dominates the transition, and hence we know the exact line
wavelength, though the singlet transition is of course weaker than the
triplet.  We therefore measured the actual observed 
wavelength difference between the
singlet and triplet helium lines for all positions where the singlet line was
easily detected, and calculated a weighted average.
The result was a difference of 1.188$\times10^{-3}\mu$m, consistent with the
triplet line being dominated by transitions to the $J=2$ state.
Therefore, the expected wavelength difference between the triplet HeI
line and the H$_2$ transition, assuming both components have the same velocity
is 9.249$\times10^{-3}\mu$m.  Observed variations from this value of course
represent different real velocity structures in the ionised and molecular gas.

Observations were also made of SAO142608, a B0 star with a visual magnitude
of $V=8.6$.  There are only very weak atmospheric absorption lines 
present in the observed wavelength range, and none of these lie near
the target emission lines.  The data shown in Figure 1 have
been crudely flux calibrated using this star, to allow relative
fluxes from the HeI and H$_2$ lines to be estimated.

\section{Results}
All data were fitted with Gaussian functions, there being no
evidence for deviations from this profile within the errors.
The actual results are shown in Figure 2, where we have plotted data for the
H$_2$ line, the HeI triplet and the relevant Br$\gamma$ data from LH96.

We have only plotted points where the signal to noise in the flux was greater
than 4.  Given our comments in the previous section we have taken the liberty
of shifting the velocity scale for the Br$\gamma$ and HeI so that they agree at
the position of peak flux.  The absolute scale adopted is the one derived from
the HeI observations.  In addition, it should be noted that the spatial
resolution of the Br$\gamma$ is less (2 arcsec pixels as opposed to the 1.5
arsec pixels for the new data presented here).

First, we compare the flux distributions of the molecular hydrogen
emission and the He I and Br$\gamma$ emission along the on-axis slit
position (top-left panel of Fig 2). As expected the He I and
Br$\gamma$ distributions trace each other very well, but the molecular
hydrogen emission peaks nearly 2 arcsec or about 0.05 pc ahead of the
ionized gas away from the exciting star. The 1-0 S(1) distribution is
also much flatter than the ionized gas in the main body of the
nebula. Similar relative distributions are seen in the off-axis
positions in Fig 2. This is consistent with the molecular hydrogen
emission arising in a thin shell just outside the ionized region which
would be predicted by both fluorescent and shock models.

In the middle panel of Fig 2 we compare the centroid velocity of the
molecular hydrogen emission along the on-axis position with that of
the He I and Br$\gamma$ line. As expected the velocity structure of
the He I is very similar to that of Br$\gamma$ since they both trace
the same volume of ionized gas. On the other hand the molecular
hydrogen does show significantly different velocity structure. There
is much less variation in the velocity along the object: only about 7
\kms\ in the molecular hydrogen compared to over 20 \kms\ for the
ionized gas. The molecular velocity structure has the appearance of an
approximately constant background emission at $v_{lsr}\approx$92 \kms\
with perturbations in the same direction as the ionized gas in the
main body of the nebula. In the region around the exciting star (2
arcsec behind the nebular peak) the ionized gas is redshifted relative
to the molecular gas by about 3 \kms\ whereas in the tail the ionized
gas gets progressively more blueshifted relative to the molecular
gas. In the head region the molecular gas again returns to the
`background' velocity whilst the ionized gas rapidly becomes more and
more blueshifted. Similar patterns are again seen in the off-axis
positions, especially in the 6 arcsec SE position where there is still
significant velocity structure in the molecular hydrogen.

The comparison of the line widths is shown in the bottom panel of Fig
2. The He I line widths again show a very similar pattern to that of
the Br$\gamma$, but are generally about 20\% narrower. The deconvolved
width is not a factor of two narrower as would be expected for purely
thermal broadening. In LH96 it was found that a turbulent velocity
component of 8.5 \kms\ was needed to explain the minimum width of the
Br$\gamma$ line and a similar value is required for the He I lines
here.  For more typical line widths seen in the HII region (24 \kms\
for HI, 20 \kms\ for HeI) and adopting the same temperature as in LH96
of 6500K, we find that a turbulent velocity of 10.5 \kms\ is required.
Turbulent velocities in the range 8--18 \kms\ can explain all of the
structure seen.

As expected the molecular hydrogen lines are much narrower although
again not as narrow as for pure thermal broadening at around 100 K
which would give a FWHM of only 0.5 \kms\ compared to the deconvolved
width of the narrowest lines of about 5 \kms. Some turbulent component
is almost certainly needed although this depends on how much of the
velocity structure is due to global flows rather than local turbulence
(see model later). The 5 \kms\ width is similar to that seen in 
NH$_{3}$ emission from hot clumps  in VLA data (Cesaroni et al.\ 1994, 1998),
and other dense molecular gas tracers such as CS
and C$^{13}$O (Churchwell, Walmsley \& Wood 1992, Cesaroni et al.\ 1992) 
in \gtn\ in larger beam radio and mm-wave
observations. As in the ionized gas there is a very rapid increase in
the line width ahead of the bow with signs of the molecular gas
returning to a more quiescent state only at the last measured position
of the on-axis data. A similar although less rapid increase in line
width appears in the tail, again very similar to the ionized gas.

\section{Discussion}

The advantage of these new data is the very accurate comparison of the
velocity structure in the ionized and molecular gas. Along the axis of
the cometary H II region the velocity difference between the ionized
and molecular gas never exceeds about 4 \kms. In the bow shock model
for this object (Van Buren \& Mac Low 1992) it is predicted that most
of the ionized gas emission should be redshifted relative to the
ambient cloud as the exciting star bores its way into the molecular
cloud and away from the observer. The maximum redshift should be the
order of the star's relative velocity moderated by the cosine of the
inclination angle or about 14 \kms\ in the Van Buren \& Mac Low
model. From our data this redshift is at most about 6 \kms.

In the tail region of \gtn\ our data show strong evidence for a
champagne flow with the ionized gas becoming progressively more
blueshifted relative to the molecular gas. Moving into the main body
of the nebula where the ionized gas is redshifted relative to
the molecular gas we can understand this motion as
expansion of the ionization front into the cloud behind (see Fig 5 of
LH96 to help visualize the geometry of the object). The small
amount of velocity structure in the molecular hydrogen emission
through this main body region can also be understood as expansion. If
the molecular hydrogen emission arises from a thin shell around the
ionized zone that is already being set in motion by the expanding
nebula then the observed velocity structure would result.

It is in the head region of the object where it becomes difficult to translate
the observed velocity structure into any simple explanation or existing
model. As one proceeds further into the head the ionized gas becomes
progressively more and more blueshifted relative to the molecular gas. There
are hints that the molecular gas begins to be dragged blue-wards as well before
decoupling from the ionized gas at about the +5 arcsec offset along the on-axis
position. Commensurate with the increasing blueshift of the ionized gas is a
similarly dramatic increase in the line width. In a homogeneous flow picture
this pattern in the line centroid and width has to be due to acceleration along
the line of sight. Inspection of the geometry for \gtn, i.e. inclined at
-135$^\circ$ (see Fig 5 of LH96), led us to explore the possibility of a
substantial component of motion in the ionized gas tangential to the shell. Due
to the orientation of the object tangential motion would be fully towards the
observer in the head region whilst being tangential to the line of sight (and
hence not observed) in the main body of the nebula. Physically this motion
could be seen as a champagne flow where instead of the gas streaming straight
along the axis of the object it is forced around the walls of the shell by the
confining action of the stellar wind. An equivalent effect can be generated in
the bow-shock model by adding gas pressure forces to the ram pressure ones they
consider.  Clearly, for lower stellar velocities, as we find here for \gtn, gas
pressure forces play a more important role.  However, in order to explain the
increasing blueshift in the ionized gas we have to increase the speed of the
tangential motion with distance from the shell. This seems to run against what
one would expect physically. The tangential motion would accelerate from zero
at the apex and so would naturally explain the large line widths along the line
of sight in the head region.

In an attempt to quantify these ideas and test the expansion
interpretation of the molecular hydrogen emission we have revamped the
empirical model of LH96. The tangential motion described above was
quantified by the form
\begin{equation}
v_{tan}=v_{tan}(max)(\phi/\pi)(l/l_{max})^{\alpha}
\end{equation}
Hence the tangential component accelerates linearly with $\phi$ from the apex
($\phi$=0) to the tail ($\phi \rightarrow \pi$). Ahead of the
bow the velocity increases with the standoff distance $l$ ($l_{max}$
is the peak of the ionized shell). In order to attempt to match the
on-axis data ahead of the bow an exponent $\alpha$ of at least two was
necessary. However, if this was applied right around the object then
the fit in the tail was very bad. Hence the exponent $\alpha$ was
modulated such that
\begin{equation}
\alpha=2(1-\phi/\pi)
\end{equation}

The model in LH96 also had a much sharper cut-off in density ahead of the bow
than observations appear to show. Both K-band, Br$\gamma$ and radio continuum
images (LH96, Watson et al. 1997, Fey et al. 1995) all show emission extending
away from the head at about the 10 per cent level for at least a couple of
arcseconds.  We suspect that there is some kind of partial ionization zone
where ionizing photons are leaking into the dense gas ahead of the cometary
region.  To account for this we added an additional Gaussian shell component to
the density distribution on the outside of the shell whose peak density was
three times lower than the main shell and whose width was ten times
thicker. This provided a much better match to the flux distribution ahead of
the bow.

Even after accounting for the partial ionization zone a purely
tangential flow did not provide a good match to the observations. In
particular the peak redshifted velocity occurred ahead of the peak flux
rather than behind as in the data. We therefore included an expansion
component as in the champagne model in LH96. The expansion
was perpendicular to the shell, i.e. orthogonal to the tangential
component and can be physically identified with expansion of the
ionization front. In order not to produce line-splitting in the tail
region the expansion velocity was modulated such that
\begin{equation}
v_{exp}=v_{exp}(max)(1-\phi/\pi)
\end{equation}

A model with $v_{tan}$(max)=20 \kms\ and $v_{exp}$(max)=20 \kms\ is compared
with the observed Br$\gamma$ flux, velocity centroid and width for both the
on-axis and 6 arcsec off-axis positions in Fig 3. The essential
features of the model can be seen in the on-axis velocity
structure. Ahead of the apex at around +5 arcsec the tangential
component means that the flow is coming towards the observer whilst
the expansion component is mostly perpendicular to the line of sight. At
the position of the star (0 arcsec) the situation is reversed and we
see the expansion of the rear shell back into the molecular cloud. In
the tail the flow becomes progressively more blueshifted as the gas
accelerates towards us. The velocity field in the model over-predicts
this blueshift, but further attempts to find a better empirical
formalism were not felt to be warranted. A significant increase in the line
width does occur in the head region for this model although still not
enough to match the observations.  The fit is still poor in the
off-axis positions. Even so, the fact that there is a turnover in the
velocity centroid is an improvement over the models in LH96.

We have developed a similar empirical model to investigate the
molecular hydrogen velocity structure. The density distribution was
assumed to have the same bow-type structure as the main ionized gas
shell, but with a standoff distance $l$ some 0.5 arcseconds
larger. Since we believe the H$_{2}$ emission is mostly fluorescent we
assumed the flux was proportional to density and the inverse square of
the distance from the exciting star to take account of the geometric
dilution of the UV radiation field. Fig 4 shows the results from such
a model compared with the observed velocity structure. This model
includes the same expansion component as did the model for the ionized
gas with $v_{exp}$(max)=10 \kms\ , but has no tangential component to
the motion. It gives a reasonable explanation for the motion in the
main body of the nebula along the axis. Once again it cannot explain
the broad widths ahead of the apex and the large amount of velocity
structure off axis. The line widths in this model consisted of a
thermal component with a temperature set at 100 K and a turbulent
component of 3 \kms . This turbulent component is significantly less
than that needed in the model of the ionized gas.

If the ionized and molecular hydrogen really do share the same type of
expansion motion it would imply in a homogeneous scenario that molecular
hydrogen arises from a thin shocked region ahead of the ionization front.  The
fact that the molecular gas is shocked does not of course mean that the
observed H$_{2}$ emission is not dominated by fluorescence from the strong UV
radiation field.  

\section{Conclusions}
It is clear from our new data that the bow shock model proposed by Van Buren
and Mac-Low (1992) is not a good match to the observations. We cannot rule out
some supersonic motion of the star and its wind through the molecular
cloud. Indeed, the velocity difference between the molecular hydrogen and the
ionised gas tracers as a whole indicates that this relative motion is limited
to less than 6~kms$^{-1}$. Therefore, this cannot explain the large
20~kms$^{-1}$ velocity changes observed along the object.  Furthermore, the
evolution of the HII region around a slowly moving star is shown by
Tenorio-Tagle, Yorke \& Bodenheimer (1979).  Clearly at late times any
champagne flow element would dominate the expansion, so that the HII region
would no longer be classified as ultracompact.

It is also equally clear that the classical champagne flow model
we considered in LH96 is not a good match.  The best fit we have
found using a simple empirical model can be interpreted as a champagne
flow model in which a powerful stellar wind influences the gas
kinematics near the exciting star(s).  Certainly the implied stellar
type(s) of the exciting star(s) is consistent with the presence of
such a wind.  Watson et al.\ (1997) found that the exciting star
should be at least 60\Msolar\ from JHK photometry (note, that the
magnitudes we reported in LH96 for the exciting star(s) were in error:
the correct values are K$_n=10.6$, H$=12.1$ and J$=14.6$, and these
are now consistent within the errors with the values reported by
Watson et al).

Our new empirical model still fails to explain all the
aspects of the observed data.  In particular, two significant areas
are left unexplained.  First, the blueshifted emission ahead of the
exciting star(s) and the fact that the lines becomes broader with
increasing distance from the star. The former is perhaps indicative of
a mass-loaded flow where clumping in the gas progressively randomizes
the flow direction as it slows down.  It is less easy to see why the
velocity centroid shifts to the blue so rapidly.  The other possible
mechanism that in principle could explain both features is scattering.
Henney (1998) describes a similar situation in which he explains
broadened [OIII] emission in the Orion blister through scattering.
Unfortunately, we cannot test this possibility with our current data.
Secondly, the off-axis positions show a very similar velocity pattern
to the on-axis one.  The discrepancy between the model and the
off-axis positions probably simply reflects the limits of our
empirical model rather than anything more complex however.

In summary then, it seems clear that the dominant mechanism
determining the kinematics in \gtn\ is a combination of a reasonably
normal champagne flow channelled into a thin shell by the stellar wind.

\section*{Acknowledgements}
The United Kingdom Infrared Telescope is operated by the Joint Astronomy Centre
on behalf of the Particle Physics and Astronomy Research Council and
we are grateful for the support given during the observations. We
would like to thank Dave van Buren for valuable discussion on the relative
merits of bow-shock and champagne flow models for this source, and Alan
Watson for highlighting the difference between our previously published
photometry for the exciting star and his own.

\parindent=0pt

\vspace*{3mm}

{\bf References}\par
\mnref{Bashkin S., Stoner J.O., 1975, Atomic Energy Levels and
      Grotrian Diagrams, North Holland Publishing Company, Amsterdam}
\mnref{Bragg, S.L., Brault, J.W., Smith, W.H., 1982, \apj, 263, 999}
\mnref{Cesaroni, R., Walmsley, C.M., Kompe, C., Churchwell, E., 1992, 
	\aaa, 253, 278}
\mnref{Cesaroni, R., Churchwell, E., Hofner, P., Walmsley, C.M., Kurtz, S.,
	1994, \aaa, 288, 903}
\mnref{Cesaroni, R., Hofner, P., Walmsley, C.M., Churchwell, E., 1998, 
	\aaa, 331, 709}
\mnref{Churchwell, E., Walmsley, C.M., Wood, D.O.S., 1992, \aaa, 253, 541}
\mnref{Dyson, J.E., Williams, R.J.R., Redman, M.P., 1995, \mn, 277, 700}
\mnref{De Pree, C.G., Rodriguez, L.F., Goss, W.M., 1995, RmxAA, 31, 39}
\mnref{Fey, A.L., Gaume, R.A., Claussen, M.J., Vrba, F.J., 1995, \apj, 453,
	308} 
\mnref{Fich M., 1993, \apjs, 86, 475}
\mnref{Henney, W.J., 1998, \apj, 503, 760}
\mnref{Hollenbach, D., Johnstone, D, Lizano, S., Shu, F., 1994, \apj,
	428, 654}
\mnref{Hofner, P, Churchwell, E., 1996, \aas, 120, 283}
\mnref{Icke V., Gatley I., Israel F. P., 1980, \apj, 236, 808}
\mnref{Israel F. P., 1978, \aaa, 90, 769}
\mnref{Kurtz S., Churchwell E., Wood D. O. S., 1994, \apjs, 91, 659}
\mnref{Lumsden, S.L., Hoare, M.G., 1996, \apj, 464, 272}
\mnref{Mac Low M.-M., Van Buren D., Wood D. O. S., Churchwell E., 1991, 
	\apj, 369, 395}
\mnref{Oliva E., Origlia L., 1992, \aaa, 254, 466}
\mnref{Redman, M.P., Williams, R.J., Dyson, J.E., 1996, \mn, 280, 661}
\mnref{Redman, M.P., Williams, R.J., Dyson, J.E., 1998, \mn, 298, 33}
\mnref{Tenorio-Tagle, G., Yorke, H.W., Bodenheimer, P., 1979, \aaa, 80, 110}
\mnref{Van Buren D., Mac Low M.-M., 1992, \apj, 394, 534}
\mnref{Watson, A.M., Coil, A.L., Shepherd, D.S., Hofner, P., Churchwell, E.,
	1997, \apj, 487, 818}
\mnref{Williams, R.J.R., Dyson, J.E., Redman, M.P., 1996, \mn, 280, 667}
\mnref{Wood D. O. S., Churchwell E., 1989, \apjs, 69, 831}
\mnref{Yorke, H.W., 1986, ARA\&A, 24, 49}

\clearpage
\onecolumn

\begin{center}

\begin{minipage}{7in}{

\psfig{file=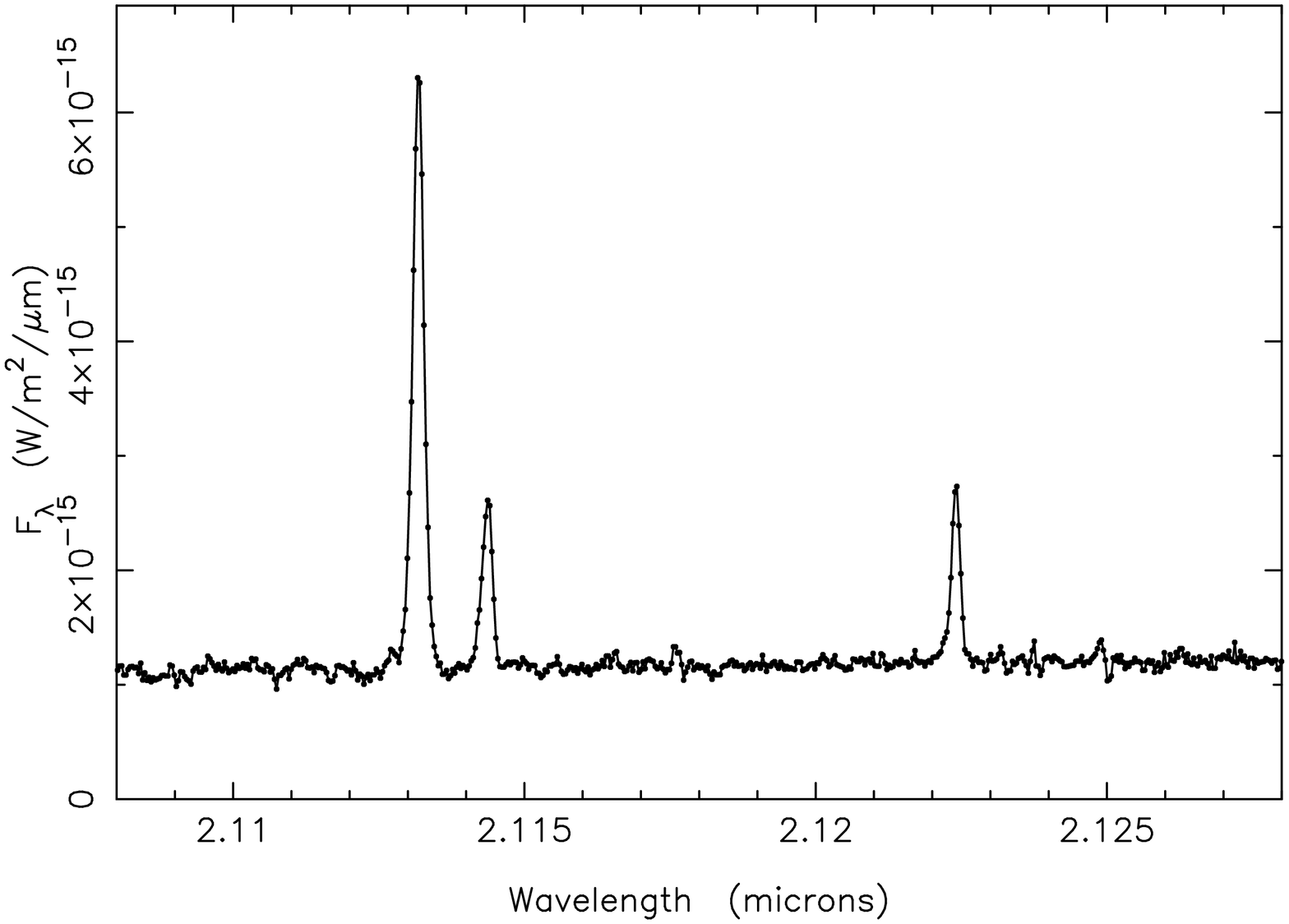,angle=-90,width=5in,bbllx=80pt,bburx=600pt,bblly=20pt,bbury=720pt,clip=}

}\end{minipage}\end{center}

\noindent{\bf Figure 1:} Extracted spectrum of \gtn\ for a 1 arcsecond by 4.5
arcsecond region around the flux peak.  The lines are (from left to right) the
4$^{3}$S--3$^{3}$P and 4$^{1}$S--3$^{1}$P transitions of HeI, and the 1--0 S(1)
transition of H$_2$.  The absolute flux calibration is only approximate.

\newpage

\begin{center}

\begin{minipage}{12in}{

\psfig{file=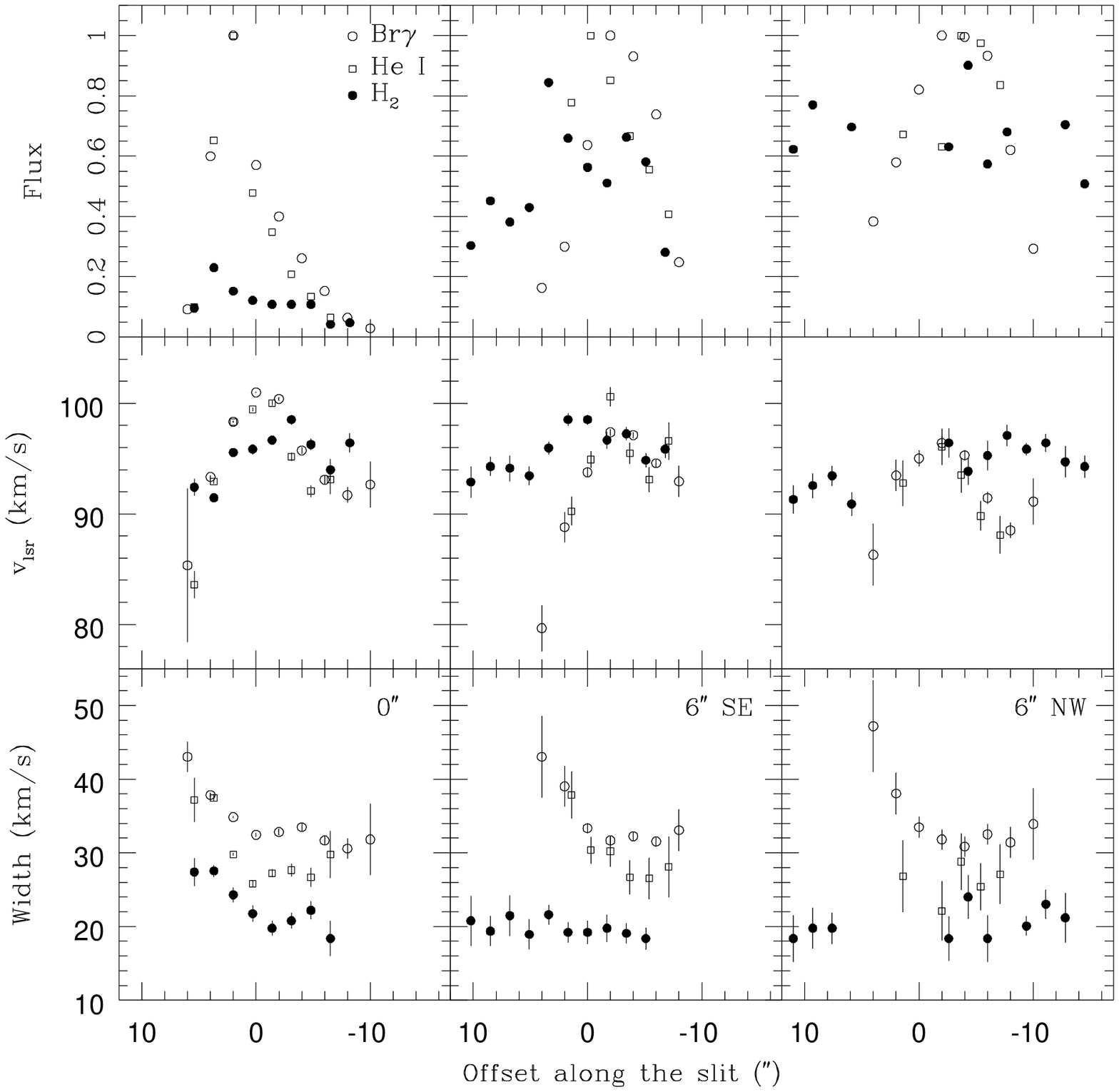,width=6.5in,bbllx=20pt,bburx=580pt,bblly=145pt,bbury=690pt,clip=}

}\end{minipage}\end{center}

\noindent{\bf Figure 2:} Comparison of the observed flux, velocity
centroid and line width along the slits for the three lines \brg\, He
I and 1-0 S(1) H$_{2}$. The three different vertical panels show the
results for the three different slit positions - on the axis of the
object, 6 arcseconds to the SE and 6 arcseconds to the NW. The \brg\ and He
I fluxes are normalized and their spatial profiles agree well as
expected. The 1-0 S(1) H$_{2}$ flux is shown relative to the \brg\
flux. The zero offset along the slit corresponds to the position of
the exciting star.

\vfill
\clearpage

\begin{center}

\begin{minipage}{12in}{

\psfig{file=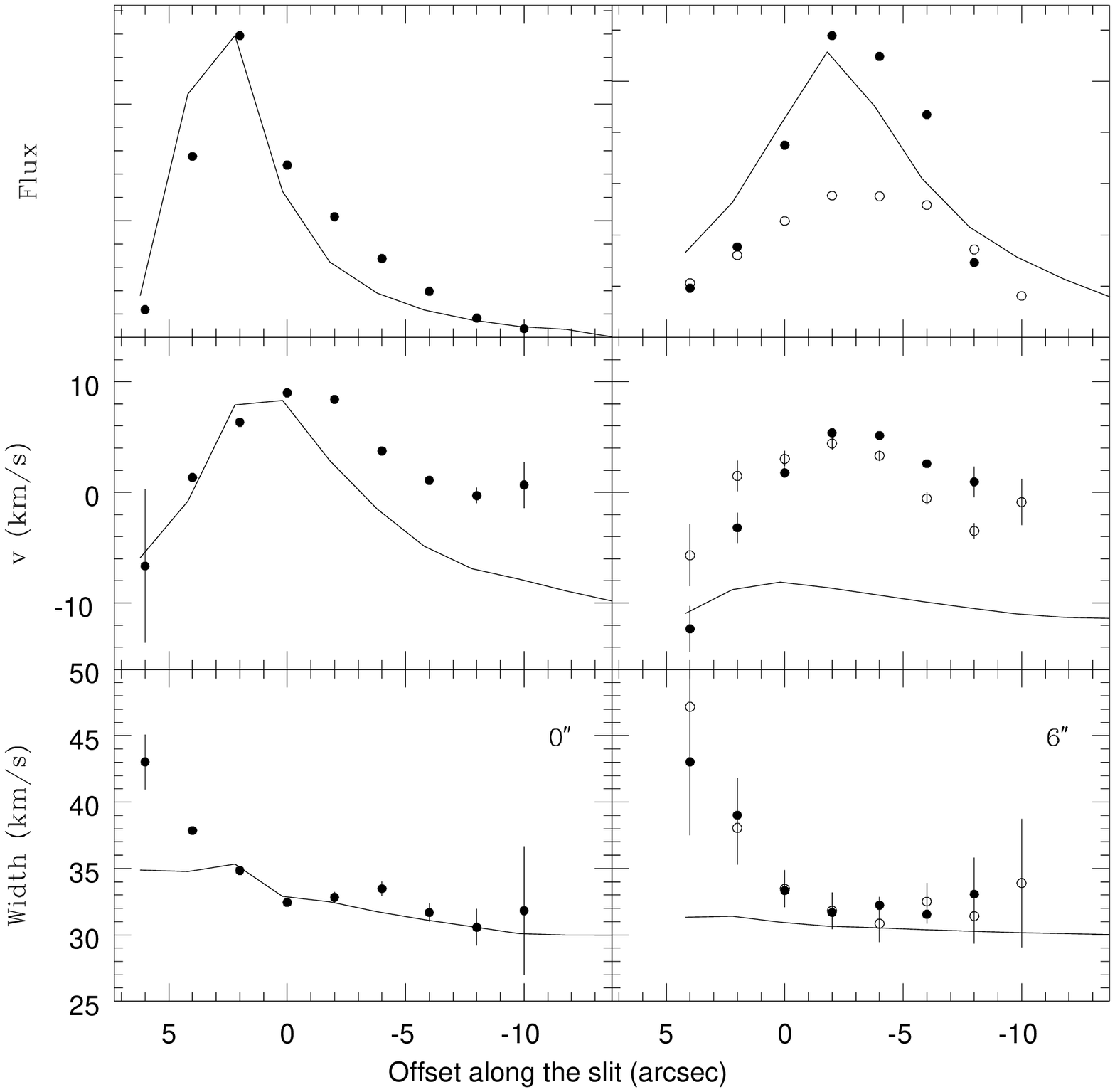,width=5in}

}\end{minipage}\end{center}

\noindent{\bf Figure 3:} Comparison of the empirical model (solid
line) for the ionized gas with the \brg\ data (points). In the right
hand panel the solid circles are the 6 arcsec SE data and the open
circles are the 6 arcsec NW data. The model flux has been normalized
to the peak flux of the on-axis data. The velocities are now with respect
to the undisturbed ISM (or star); a $v_{lsr}$=92 kms$^{-1}$ has
been subtracted from the data.

\vfill
\clearpage

\begin{center}

\begin{minipage}{12in}{

\psfig{file=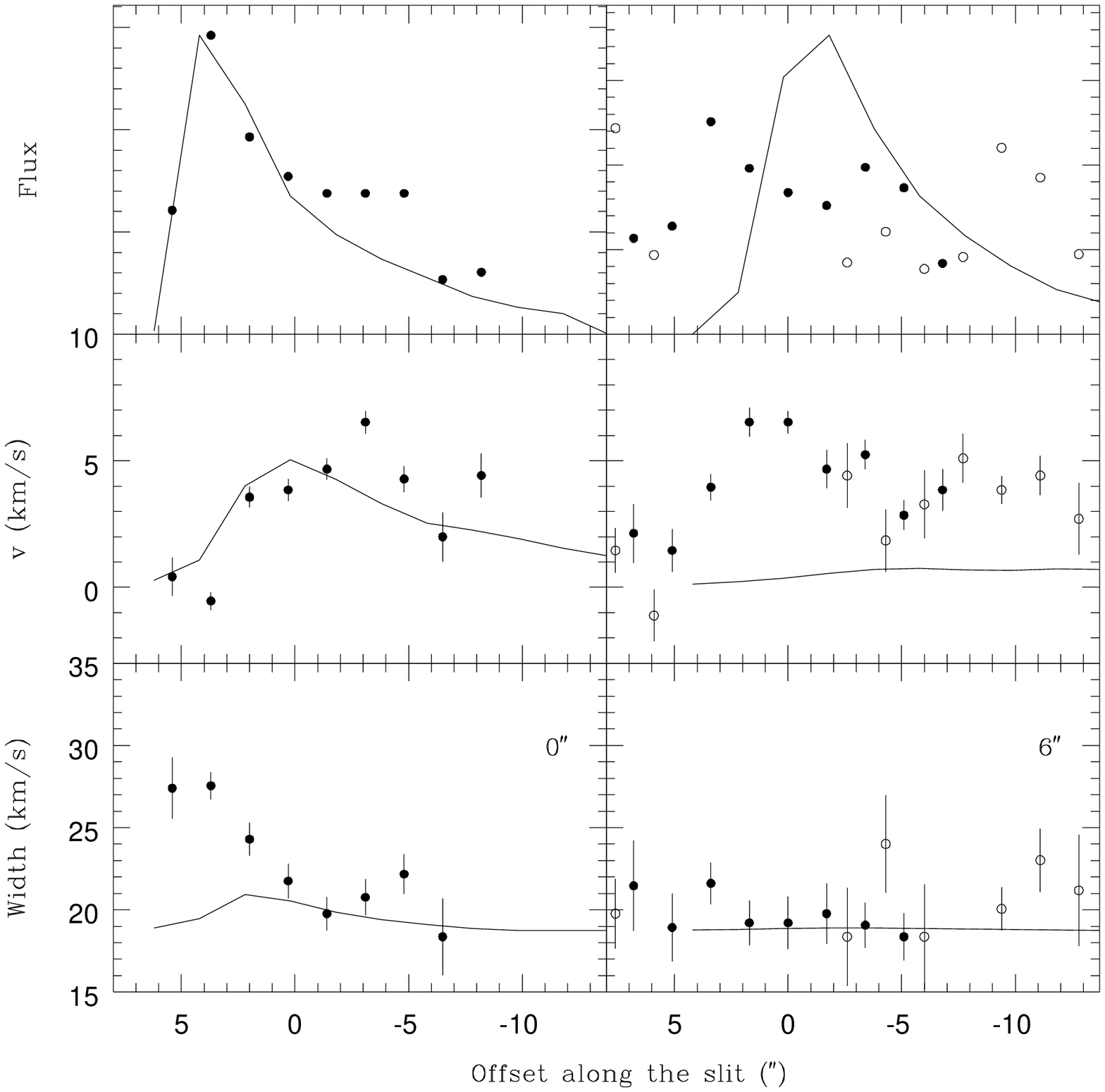,width=5in}

}\end{minipage}\end{center}

\noindent{\bf Figure 4:} Comparison of the empirical model (solid
line) for the molecular gas with the H$_{2}$ data (points). A
$v_{lsr}$=92 kms$^{-1}$ has also been subtracted from the observed velocities.

\end{document}